\documentclass[11pt,twoside]{article}
\usepackage{newpasp}

\usepackage{epsf}
\usepackage{psfig}

\index{Begelman, M. C.}

\begin{document}

\title{Impact of AGNs on the Surrounding Medium}
\author{Mitchell C. Begelman}
\affil{JILA, University of Colorado, Boulder, CO 80309-0440}

\begin{abstract}
The time- and ensemble-averaged mechanical energy outputs of radio galaxies may be large enough to offset much of the cooling inferred from X-ray observations of galaxy clusters.  But does this heating actually counterbalance the cooling, diminishing cooling flows or quenching them altogether?  I will argue that energy injection by radio galaxies may be important even in clusters where no active source is present, due to the likely intermittency of the jets.  If the energy injected by radio galaxies percolates through the intracluster medium without excessive mixing, it could stabilize the atomic cooling responsible for X-ray emission.

\end{abstract}

\section{Introduction}
The impact of radio galaxies on their surroundings is probably far out of proportion to their numbers.  Although only a small minority of AGNs seems to produce powerful jets, virtually all of the mechnical energy output is transferred to the ambient medium.  Relatively little of this energy can be radiated away (Scheuer 1974), and therefore the majority must go into heat and motion.  In contrast, the large, easily detected supplies of radiant energy pouring out of most AGNs probably have little effect on the surroundings. Cool, dense clouds ($\la 10^4$ K) in the interstellar medium of the host galaxy can easily reradiate this energy, while the hotter, diffuse gas responds thermally through the Compton effect, with a low efficiency $\sim \tau (h\nu/m_e c^2)$, where $\tau\ll 1$ is the electron scattering optical depth.  Moreover, with Compton temperatures $\la 10^7$ K for typical AGN spectra, irradiation is more likely to {\it cool} the hot phase of the ISM/ICM through the inverse Compton effect, rather than heat it. 

Nevertheless, in a time-averaged or ensemble-averaged sense, there seems to be plenty of energy associated with radio jets alone.  Peres et al.~(1998) compared the radio and X-ray luminosities of 58 clusters in a {\it ROSAT} 
flux-limited survey.  The ratio of 1.4 GHz power to bolometric X-ray power from within the ``cooling radius" of those sources identified as cooling flows was $\sim 0.008$, just shy of 1\%.  Now, the maximum possible synchrotron emissivity of radio lobes in internal pressure equilibrium is typically smaller than a few percent of the kinetic power, assuming equipartition and unit filling factor of relativistic electrons in the lobes.  This fraction evolves toward smaller values as the source expands (Begelman 1996, 1999), and is also decreased by any deviation from equipartition. (Note that it can be increased somewhat if the emission comes mainly from small regions at high pressure, e.g., strong shocks within the lobes, but this seems unlikely to produce a large correction.)  Thus it is very likely that the 
ensemble-averaged jet  power is at least of the same order as the observed X-ray power, and may be larger. But one must address at least two issues before leaping to the conclusion that energy injection by radio jets offsets X-ray cooling. First, most of the radio flux from the {\it ROSAT} sample is contributed by a small number of clusters with very powerful radio sources.  Typically, only about 10\% of the cooling flow clusters show strong {\it current} activity. Second, it is notoriously difficult for simple heating mechanisms to balance X-ray cooling in a stable way.  I will attempt to address these issues below.

\section{Are Radio Galaxies Intermittent?}

There are several pieces of circumstantial evidence that point to intermittency in radio galaxies. At a morphological level, low-frequency images such as the 90 cm map of M87 shown by Owen at this meeting (Owen et al. 2000) sometimes show large, diffuse regions of radio emission that appear distinct from (and much older than) the currently active radio lobes. The intermediate scale structures apparent on the M87 image appear consistent with the early rise of buoyant lobes in a ``dead" source (Churazov et al. 2000a,b; Reynolds et al. 2000).  

That we can see these diffuse relics in M87 and a few other sources may be our good fortune. Simple models for the evolution of radio galaxies once the jets have turned off suggest that they fade very rapidly, due both to adiabatic expansion of the relativistic electrons and to decreasing magnetic field strengths (Reynolds \& Begelman 1997). When these faint haloes are detected, it is generally through low-frequency radio observations.  They might conceivably be detectable as well in  X-rays.

Finally, there may be statistical evidence for intermittency embedded in
the number-vs.-size counts of small luminous radio galaxies. The ``plateau"
in source counts noted by O'Dea \& Baum (1997) can be explained if young
radio galaxies switch on and off with a duty cycle of $\sim 0.1$ and a time
scale of $\sim 10^5-10^6$ yr (Reynolds \& Begelman 1997; Fig. 1).
\begin{figure}
\psfig{figure=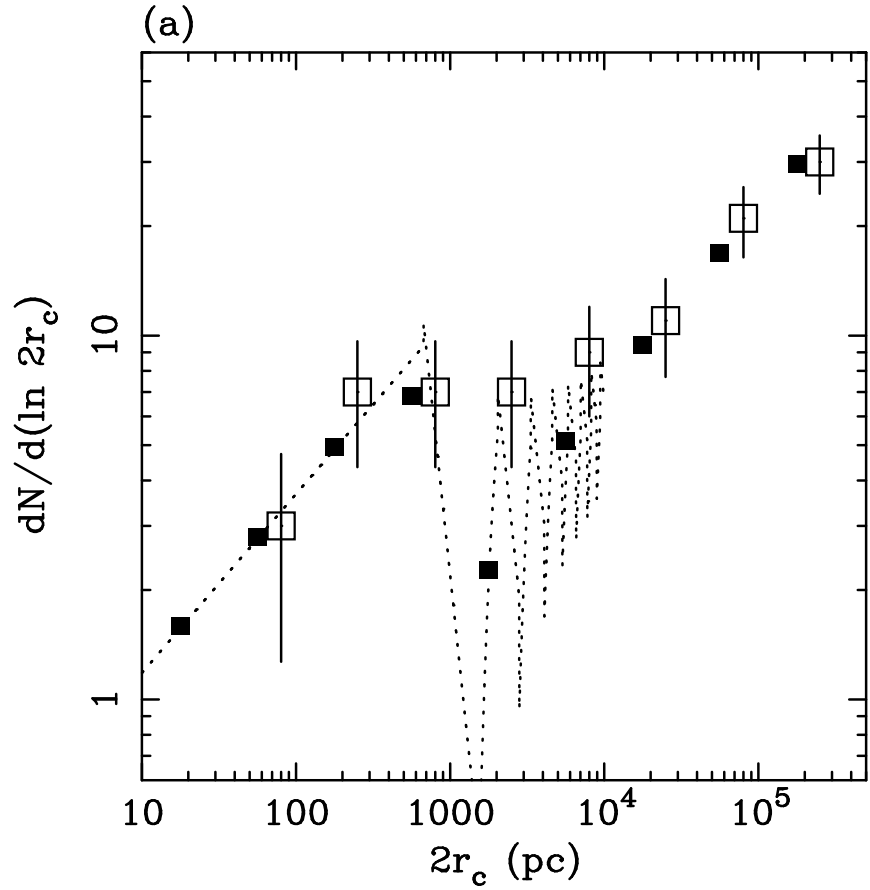,width=0.95\textwidth,angle=270}
\caption{\small Open squares show number-vs.-size counts of radio galaxies from O'Dea \& Baum (1997).  The ``plateau" between 300 and 3000 pc can be explained by source intermittency. The dashed line shows a typical theoretical model, while filled squares show a binned theoretical distribution. See Reynolds \& Begelman (1997) for details. }
\label{fig1}
\end{figure}

\section{``Effervescent" Heating of the ICM}

It has proven notoriously difficult to find a heating mechanism that can offset X-ray cooling locally, while maintaining the kind of entropy or temperature gradient that appears to be present in X-ray emitting clusters.  Thermal conduction using standard formulae either wipes out the gradients altogether or has little effect (Bertschinger \& Meiksin 1986); models invoking cosmic-ray diffusion suffer similar problems (Loewenstein et al. 1991).  An alternative approach has been to suppose the heating to be impulsive, and the cooling episodic (Binney \& Tabor 1995); there is some suggestion of this in our numerical simulations of evolving radio galaxy cocoons, which show a large fraction of the power going into gravitational potential energy of ICM being ``lifted" by the buoyant cocoon (Reynolds et al. 2000).  

I would like to propose another, more gentle mode of energy transfer, which I call ``effervescent" heating.  Suppose the radio source deposits some buoyant form of energy --- relativistic particles or bouyant gas --- which distributes itself  relatively evenly among bubbles or filaments but does not mix microscopically with the ICM.  These bubbles will then rise through the ICM, like bubbles in a glass of champagne.  Because of the non-negligible pressure gradient, the bubbles (or filaments) will expand as they rise, doing $pdV$ work on their surroundings. The expansion, of couse, converts internal energy to kinetic form, but it is likely that the resulting disorganized motion of the ICM is quickly converted to heat.  In a steady state (and assuming spherical symmetry), the energy flux available for heating is 
\begin{equation}
\dot E \propto p_b(r) ^{(\gamma-1)/\gamma }
\end{equation}
where $p_b(r)$ is the partial pressure of buoyant fluid inside the bubbles at radius $r$ and $\gamma$ is the adiabatic index of the buoyant fluid. If we can determine the partial pressure --- let's suppose it scales with the thermal pressure in the ICM, for the sake of argument --- then one can derive a volume heating function with some interesting properties:
\begin{equation}
{\cal H } \sim - \nabla \cdot {\dot E \over 4\pi r^2} \propto {p^\beta \over r^3}{d\ln p \over d\ln r} 
\end{equation}
where $\beta = (\gamma - 1)/ \gamma $.  The parameter $\beta$ ranges between 1/4 for an ultrarelativistic fluid and 2/5 for nonrelativistic thermal gas. Unlike heating functions that depend on local microscopic physics, this heating function is nonlocal in the sense that it depends on the pressure gradient, i.e., on the hydrostatic structure of the cluster atmosphere.  In this regard, it  resembles the kind of heating function one gets from thermal conduction, but here the heating rate depends on the gradient of pressure rather than temperature. This feature yields just the sort of feedback necessary to stabilize X-ray cooling, which has the form ${\cal C} = n^2 \Lambda(T)$. Suppose the cluster gas has managed to find a thermal equilibrium between effervescent heating and cooling.  If this is perturbed so that cooling is excessive, then the pressure gradient will steepen, increasing the intensity of the heating and bringing the atmosphere back towards equilibrium.  In this case, there would be no cooling flow, although there would be copious X-ray cooling.

One can derive a crude global stability criterion under the assumption of a singular isothermal cluster potential and linear perturbations away from an isothermal atmosphere. If we define a ``cooling index" $\alpha \equiv d\ln \Lambda / d\ln T$, we find that X-ray cooling can be stabilized for 
\begin{equation}
\alpha < {1-2\beta  \over 2-\beta },
\end{equation}  
as illustrated in Fig. 2. 
\begin{figure}
\psfig{figure=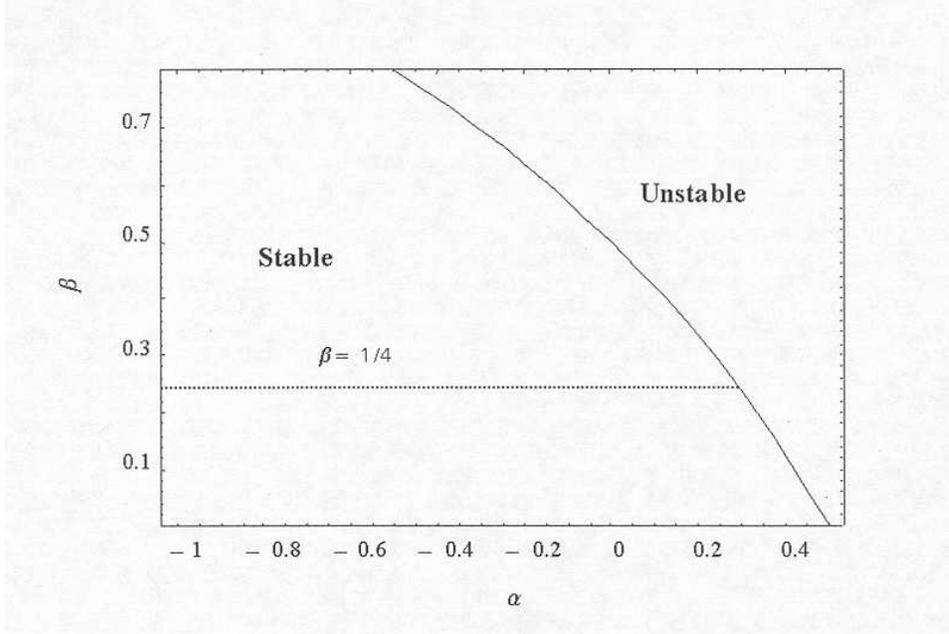,width=0.95\textwidth}
\caption{\small Domains of stability and instability for effervescent heating are separated by the solid curve.  The dotted line indicates the condition $\beta = 1/4$, corresponding to bubbles filled with relativistic ($\gamma = 4/3$) fluid.}
\label{fig2}
\end{figure}
This upper limit varies between 2/7 for relativistic bubbles ($\beta = 1/4$) and 1/8 for $\beta = 2/5$,  and always lies  below the value $\alpha = 1/2$ associated with pure bremsstrahlung. Thus bremsstrahlung cannot be stabilized by this mechanism.  However, in most regions of cluster cooling flows, the temperature lies in a range where atomic cooling dominates over free-free, and $\alpha$ is considerably smaller than 1/2 (Sarazin \& White 1987; Fig. 3).

\begin{figure}
\hbox{
\psfig{figure=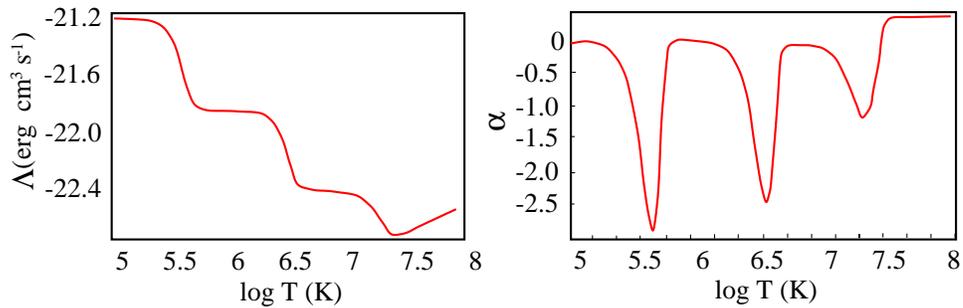,width=0.95\textwidth}
}
\caption{\small Left: Cooling function $\Lambda(T)$ for intracluster gas with solar abundances. Right: Cooling index $\alpha \equiv d\ln \Lambda / d\ln T$ as a function of temperature.}
\end{figure}

Note that the joint conditions of hydrostatic and thermal equilibrium predict that the specific entropy of the ICM should decrease strongly toward the center of the cluster, thus mimicking one of the main features of a cooling flow. This result is insensitive to the value of $\alpha$. The density scales with radius as 
\begin{equation}
\rho \propto r^{- 3/ (2 - \beta)},
\end{equation}
which is consistent with typical density profiles deduced observationally.

\section{Conclusions}
In a time- and ensemble-averaged sense, it appears that radio galaxies can supply enough energy to offset the observed cooling of intracluster gas.  But the relative rarity of active radio sources would require that they be intermittent, and that the heating effects due to mechanical energy injection persist long after the radio lobes have faded. There is good circumstantial evidence that the former is true, but whether the latter occurs is an open question.

I have speculated on a possible mechanism by which mechanical heating can offset cooling in a stable way. In order for effervescent heating to work, the buoyant fluid injected by the radio galaxy would have to spead its energy evenly through the cluster atmosphere, without mixing into the background at a microscopic level.  Intracluster ``weather" and perhaps a certain amount of thermal conduction could help with this.

\acknowledgments

My work on radio galaxies is supported by NSF grants AST-9876887 and AST-9529175.  I am grateful to my collaborators Chris Reynolds and Sebastian Heinz for their numerous insights and contributions.


\begin{references}
{\small
\reference Begelman M. C. 1996, in Cygnus A--Study of a Radio Galaxy, ed. C. Carilli and D. Harris, 209 (Cambridge: Cambridge Univ. Press)
\reference Begelman M. C. 1999, in The Most Distant Radio Galaxies, ed. H. J. A. R\"ottgering, P. N. Best, and M. D. Lehnert, 173 (Amsterdam: Royal Netherlands Acad. of Arts and Sciences)
\reference Bertschinger E., Meiksin A. 1986, ApJ, 306, L1
\reference Binney J., Tabor G. 1995, MNRAS, 276, 663
\reference Churazov E., Forman W., Jones C., B\"ohringer H. 2000a, A\&A, 356, 788
\reference Churazov E., Br\"uggen M., Kaiser C. R., B\"ohringer H., Forman W. 2000b, astro-ph/0008215
\reference Loewenstein M., Zweibel E. G., Begelman M. C. 1991, ApJ, 377, 392 
\reference O'Dea C. P., Baum S. A. 1997, AJ, 113, 148
\reference Owen F. N., Eilek, J. A., Kassim N. E. 2000, ApJ, submitted (astro-ph/0006150)
\reference Peres C. B., Fabian A. C., Edge A. C., Allen S. W., Johnstone R. M., White D. A. 1998, MNRAS, 298, 416
\reference Reynolds C. R., Begelman M. C. 1997, ApJ, 487, L135
\reference Reynolds C. R., Heinz S., Begelman M. C. 2000, ApJ, submitted
\reference Sarazin C. L., White R. E. 1987, ApJ, 320, 32 
\reference Scheuer P. A. G. 1974, MNRAS, 166, 513
}
\end{references}
\end{document}